\documentclass[twocolumn,superscriptaddress,showkeys]{revtex4-1}
\usepackage[T1]{fontenc}
\usepackage[utf8]{inputenc}
\usepackage{float}
\usepackage{graphicx}

\usepackage{amssymb}

\begin{document}

\title{\vspace{-15mm}\fontsize{20pt}{10pt}\selectfont\textbf{Important role of projectile excitation in $^{16}$O+$^{60}$Ni and $^{16}$O+$^{27}$Al scattering at intermediate energies}}

\author{V.A.B. Zagatto}
\email{Email address: vinicius.zagatto@gmail.com}
\affiliation{{\it Instituto de F\'isica da Universidade Federal Fluminense, 24210-346, Niter\'oi, RJ, Brazil}}
\author{F. Cappuzzello}
\affiliation{{\it Istituto Nazionali di Fisica Nucleare - Laboratori Nazionali del Sud, 95123, Catania, CT, Italy}}
\affiliation{{\it Dipartimento di Fisica e Astronomia Universit\`a di Catania, I-95123, Catania, CT, Italy}}
\author{J. Lubian}
\affiliation{{\it Instituto de F\'isica da Universidade Federal Fluminense, 24210-346, Niter\'oi, RJ, Brazil}}
\author{M. Cavallaro}
\affiliation{{\it Istituto Nazionali di Fisica Nucleare - Laboratori Nazionali del Sud, 95123, Catania, CT, Italy}}
\author{R. Linares}
\affiliation{{\it Instituto de F\'isica da Universidade Federal Fluminense, 24210-346, Niter\'oi, RJ, Brazil}}
\author{D. Carbone}
\affiliation{{\it Istituto Nazionali di Fisica Nucleare - Laboratori Nazionali del Sud, 95123, Catania, CT, Italy}}
\author{C. Agodi}
\affiliation{{\it Istituto Nazionali di Fisica Nucleare - Laboratori Nazionali del Sud, 95123, Catania, CT, Italy}}
\author{A. Foti}
\affiliation{{\it Istituto Nazionali di Fisica Nucleare - Laboratori Nazionali del Sud, 95123, Catania, CT, Italy}}
\author{S. Tudisco}
\affiliation{{\it Istituto Nazionali di Fisica Nucleare - Laboratori Nazionali del Sud, 95123, Catania, CT, Italy}}
\author{J.S. Wang}
\affiliation{{\it Institute of Modern Physics, Chinese Academy of Sciences, Lanzhou, 730000, China}}
\author{J.R.B. Oliveira}
\affiliation{{\it Instituto de F\'isica da Universidade de S\~ao Paulo, 05508-090, S\~ao Paulo, SP, Brazil}}
\author{M.S. Hussein} 
\affiliation{{\it Instituto de F\'isica da Universidade de S\~ao Paulo, 05508-090, S\~ao Paulo, SP, Brazil}}
\affiliation{{\it Instituto Tecn\'ologico de Aeron\'autica, DCTA, 12.228-900 S\~ao Jos\'e dos Campos, SP, Brazil}}
\affiliation{{\it Instituto de Estudos Avan\c{c}ados, Universidade de S\~ao Paulo C.P. 72012, 05508-970 S\~ao Paulo-SP, Brazil}}

\begin{abstract}
The elastic scattering angular distribution of the $^{16}$O$+^{60}$Ni
system at $260$ MeV was measured in the range of the Rutherford cross section down to $7$ orders of magnitude below. The cross sections of the lowest $2^{+}$ and $3^{-}$ inelastic states of the target were also measured over a several orders of magnitude range. Coupled channel (CC) calculations were performed and are shown to be compatible with the whole set of data only when including the excitation of the projectile and when the deformations of the imaginary part of the nuclear optical potential are taken into account. Similar results were obtained when the procedure is applied to the existing data on $^{16}$O$+^{27}$Al elastic and inelastic scattering at $100$ and $280$ MeV. An analysis in terms of Dynamical Polarization Potentials (DPP) indicate the major role of coupled channel effects in the overlapping surface region of the colliding nuclei.

\end{abstract}


\maketitle

\section{Introduction}

The elastic scattering is usually the simplest and most intense channel in nuclear collisions. Notwithstanding, studies on it are commonly performed aiming to investigate the nucleus-nucleus potential or the effect of non-elastic channels. There are many examples in the literature where a detailed description of elastic scattering is essential to understand emerging phenomena, like the threshold anomaly on tighly \cite{ta1,ta2,ta3} and weakly \cite{tahuss,ta4,ta5,ta6} bound nuclei, the break-up dynamics \cite{kundu,zagatto,vasi,apak} and the nuclear rainbow \cite{gold1,gold2,gold3,khoa}.      
	
Most of the above mentioned studies are performed at energies around the Coulomb barrier, where the number of open channels is relatively small. Even in this region, coupling channel calculations including all the open channels are challenging, as demonstrated in \cite{canto1,keeley1,keeley2,back,canto2}. At higher incident energies, they become even more difficult as more channels are energetically open.

Recent advances in experimental techniques allow the measurement of high precision and accurate data down to very low absolute cross sections ($\sim$ nb/sr). The increased resolution reveals new phenomena \cite{kisamori,pereira}, and provide information on channel couplings that could not be studied in details before \cite{gr,capu}. From the theoretical side, the main challenge is to incorporate, in a simple and accurate model, the complex picture arising from the full many-body scattering problem. In particular, a reaction model that uses a global potential (which captures the average features of the scattering) and requires few relevant reaction channels is highly desirable.       

In Refs. \cite{ohkubo} and \cite{linares}, direct reaction calculations using global potentials have already been performed, coupling only a few relevant reaction channels. The first work demonstrated that the $3^{-}$ state of the $^{16}$O
nucleus is crucial for the appearence of a secondary rainbow structure in the
elastic scattering of $^{16}$O$+^{12}$C reaction at $330$ MeV. The effect of couplings to $^{16}$O states was not fully explored in the analysis of $^{16}$O$+^{27}$Al system (at $280$ MeV) in Ref. \cite{linares}. Studying the possible effect of projectile excitation on a third system would help to elucidate this question. In the present work, we carefully investigate one more reaction, $^{16}$O + $^{60}$Ni at E= 260 MeV to better understand the role of the excitation of the projectile.   
We chose this system as new high precision
experimental data are available on both, the elastic and inelastic scatterings.

In the next section, some details of the experiment are presented. The theoretical analysis is presented in section \ref{sec:Theoretical-Calculations}. In section \ref{sec:discussion}, discussions about the normalization of the imaginary potentials and the coupling effects of each channel on the effective potential are made. Finally, in the last section the conclusions are presented.

\section{Experimental Details}

The experimental data have been measured at the MAGNEX large acceptance
spectrometer \cite{cunsolo1,magnex} at INFN-LNS (Catania, Italy).
The $^{16}$O$^{8+}$ beam impinged on isotopically enriched $^{60}$Ni
targets. A $150$ $\mu$g/cm$^{2}$ thick target was used for
measurements at forward angles ( $5^{\circ}<\theta_{LAB}<15^{\circ}$), and a $500$
$\mu$g/cm$^{2}$ thick one was used for the large angular region ($8^{\circ}<\theta_{LAB}<45^{\circ}$).
The scattered ejectiles were momentum selected by the spectrometer
and the trajectory parameters (i.e. position, incident angles, energies)
were measured by the Focal Plane Detector (FPD) \cite{manuela}.
Particle identication and high order trajectory reconstruction techniques
allowed to recover the scattered energies and angles at the target
position \cite{cappuzzello2,manuela2}. Details of $^{16}$O$+^{27}$Al
data reduction can be found in Refs. \cite{manuela3,cappuzzello3}.

A typical energy spectrum of the $^{16}$O$+^{60}$Ni reaction is
presented in Fig. $1$. The energy resolution is about $600$
keV. The $0^{+}$ ground state, the $2^{+}$ ($E^{*}=1.33$ MeV) and
$3^{-}$ ($E^{*}=4.04$ MeV) excited states of target were clearly
resolved from each other. It is also noticed the presence of a large
structure in the $5$ to $9$ MeV region, composed by several states.
The same structure appears in $^{60}$Ni($\alpha$,$\alpha^{'}$)
scattering data of Ref. \cite{youngblood}. In this energy range,
numerous excited states of projectile ($0(2)^{+}$, $3^{-}$, $2^{+}$,
$1^{-}$, $2^{-}$ ) and target are expected to be populated. 

The angular
distributions of the absolute elastic and inelastic cross-sections were extracted
from measured yields down to a remarkable value of $10^{-7}$ $\sigma_{Ruth}$
and are shown in Fig. 2. A systematic error in the cross section is common to all data points and it is not included in the error bars. This kind of error comes from two sources: the determination of the target thickness (obtained previously measuring the energy loss of alpha particles); the beam integration by the Faraday cup (small noise induced in the Faraday cup cabling). The uncertainties represented in Fig.2 come from other sources, such as the solid angle determination (correlated to the error in the measurement of the central angular position of MAGNEX spectrometer) and the statistical error (greater at backward angles). No background subtraction was performed. A possible contribution from contaminants in the $^{60}$Ni target (usually from carbon or oxygen build-up during beam exposure) is negligible. At very forward angles ($\theta_{c.m.}<7.5^{\circ}$), the firt inelastic state of $^{60}$Ni slightly interferes with the elastic peak due to the low resolution of the detector at this angular region. See Supplemental Material at \cite{suppl} to access the data tables. For further information about the uncertainty propagation, one may consult \cite{manuela2}.     

The large uncertainty in the measured elastic data in the region of $\theta_{c.m.}>30^{\circ}$ do not allow to define a fine oscillatory pattern of scattering as done previously for the $^{16}$O$+^{27}$Al reaction at $280$ MeV (\citep{linares}) due to the low number of counts in each angle.

\begin{figure}[H]
\begin{centering}
\includegraphics[width=0.5\textwidth]{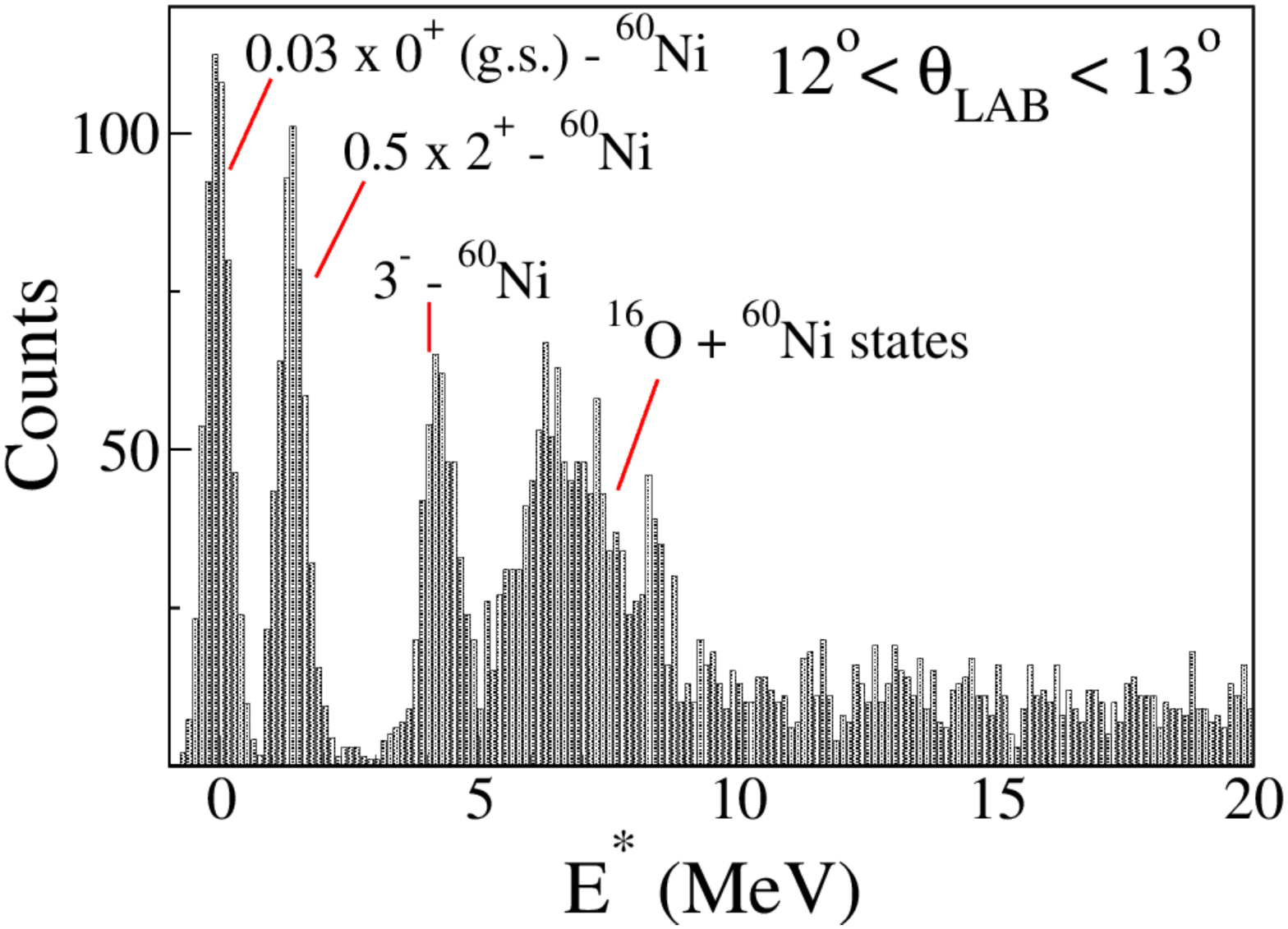} \caption{\textsl{\small{}{}Color online - Energy spectrum for the $^{16}$O$+^{60}$Ni reaction at $260$ MeV between $12^{\circ}$ and $13^{\circ}$ (laboratory framework). The most relevant peaks are identified in the figure. Note that peaks corresponding to the elastic and first inelastic states are scaled for a better visualization.}}
\par\end{centering}
\label{espectro} 
\end{figure}

\section{Theoretical Framework \label{sec:Theoretical-Calculations}}

As stated before, at low energies only a few reaction mechanisms are important, being usually included in the coupling scheme of CC calculations. Thus, the only imaginary potential needed is a internal potential to the barrier, that accounts for the loss of flux to fusion channel. When the energy increases, the number of direct reaction channels increases rapidly and it is almost impossible to couple all of them in CC calculations. This is especially true in the case of broad resonances (giant resonances) and high excited  states, for which there is no spectroscopic information. Therefore the CC calculation is performed using explicit coupling of the elastic channel to several inelastic channels corresponding to the excitation of collective states where spectroscopic information is available, and account for the coupling to the other high energy collective channels implicitly through the use of a deformed complex coupling potential. Optical model and CC calculations were performed in order to analyze
the data using the FRESCO code \cite{fresco}.

\subsection{The $^{16}$O$+^{60}$Ni System }

In the $^{16}$O$+^{60}$Ni
calculations, the $2^{+}$ and $3^{-}$
excited states of target were included in the coupling scheme.
The $3^{-}$ ($E^{*}=6.13$ MeV) excitation of projectile was also
considered. The real nuclear potential used in the calculations was the double folding São Paulo potential (SPP), $V_{SPP}$ \citep{candido,chamon1,chamon2}.
In Ref. \citep{marcos},  it was demonstrated that elastic scattering of several systems over a wide energy range is well described using "bare" (no couplings are considred) optical potential defined as 
\begin{equation}
U_{bare} = (1+0.8\cdot i)V_{SPP}
\end{equation}

The SPP potential is the non-local version of the double-folding interaction. The form used in the analysis of the data is the local, energy-dependent, equivalent potential, See Refs. \cite{candido, chamon1, chamon2}. The numerical factor 0.8 in the imaginary part of the bare potential above was fixed by adjusting a large set of data, which accounts for flux loss to other channels not explicitally included in the coupling scheme, as excited states and single nucleon knockouts \cite{hussein1}. The results considering such systematics for $^{16}$O$+^{60}$Ni and $^{16}$O$+^{27}$Al reactions are, from now on, represented by a dot-dashed magenta line.

In Fig. $2$ it is possible to notice that systematics gives a poor description of data for angles larger than $20^{\circ}$. One should remember that this systematics was originally
obtained for energies close to the Coulomb barrier, including data of cross sections (normalized by the Rutherford one) around $10^{-3}-10^{-4}$. A further improvement consisted in the coupling of the above
mentioned collective inelastic channels keeping the same bare potential. From now on, calculations considering the coupling of the inelastic channels of the target  will be represented by a green dashed line. From 
calculations of $^{16}$O$+^{27}$Al reaction at $280$ MeV reported in Ref. \cite{linares},
it was noticed that the simple inclusion of inelastic channels  was
not sufficient to reproduce elastic scattering data at large
angles. The coupling of $\alpha$ transfer channel (the one with most
favoured Q value) using realistic spectroscopic amplitudes did not
modify the results. It was also discovered that the phase of the Fresnel
oscillations in the theoretical results and the experimental data was
not the same, clearly implying that additional reaction channels are required to be coupled in calculations in order to reproduce the full angular range covered by experimental data.

In Ref. \cite{satchler}, it is shown that the deformation of the
imaginary part of the nuclear potential (when considering the coupling of the inelastic channels)
can play an important role in the $^{182}$W$(d,d^{'})^{182}$W reaction. This result is in accordance with the Bohr-Mottelson unified model \citep{bm1,bm3}. Refs. \citep{satchler2,ismail} claimed that, even for light ions,
the inclusion of the deformation of the imaginary part may provide evidence that
there is a strong interference between the Coulomb and nuclear contributions,
especially when collective excitations of the nuclei are present. Giant Resonances
(GR) can, in principle, act as important doorway states for nuclear
scattering, since they represent small amplitude oscillations of nuclei.
Very few collective degrees of freedom are involved in GR and consequently
their overlap to nuclear ground state can be large. However, not much
is known about the effect of the GR on the elastic scattering. At the
bombarding energies discussed here, high-lying states of nuclei can
be accessed (including giant resonances), redistributing part of the flux
from the low lying excited states. A detailed description of the coupling
to all GR would require several parameters, as multipolarities,
centroids and widths. Instead, in the present calculations, a different
approach is proposed, by deforming of the imaginary part of the nuclear
potential.

\begin{figure*}[ht]
\begin{center}
\includegraphics[width=1.05\textwidth]{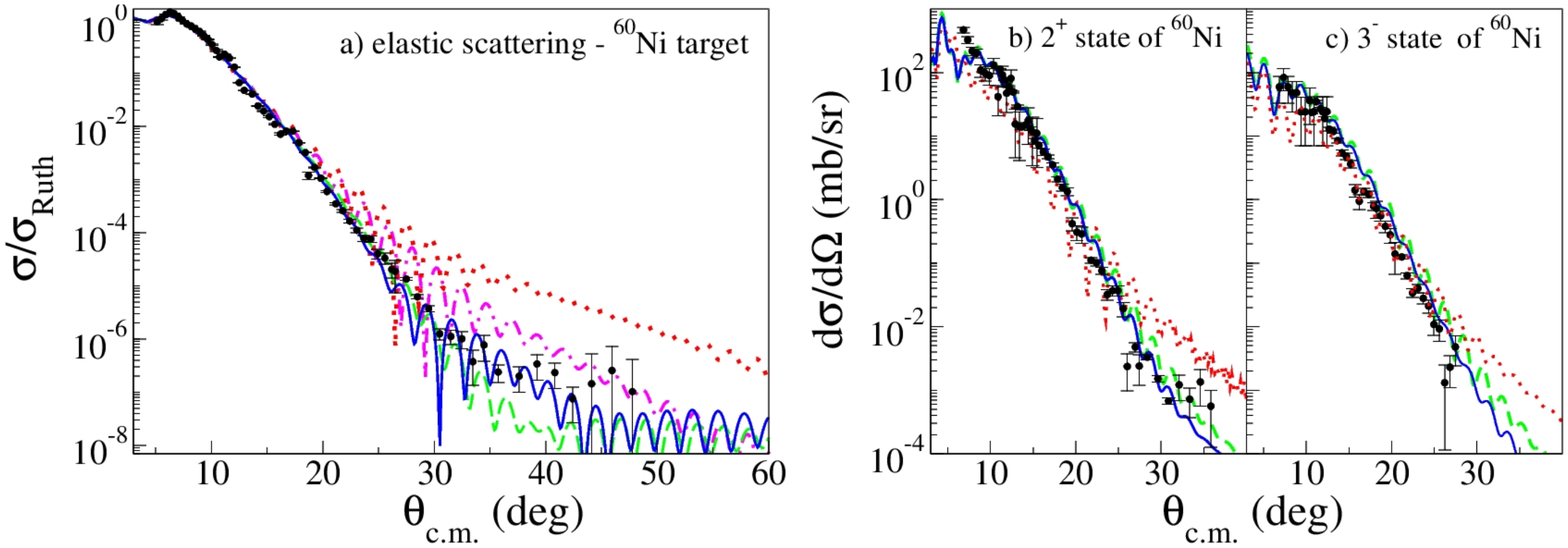} \caption{\textsl{\small{}{}(Color online) - Angular distributions of $^{16}$O$+^{60}$Ni: (a) Elastic scattering showing different calculations. The dotted-dashed magenta line shows the optical model calculation with nuclear S\~ao Paulo potential. The dashed green line shows the CC calculations including the first $2^+$ and $3^-$ inelastic states of target. The solid blue line shows the results of CC calculation including the target excitation and also the first inelastic state of projectile. The dotted red line shows the calculations including all the mentioned couplings but disregarding the deformation of the imaginary potential; (b) $2^+$ inelastic distribution of target. The calculations follow the same color representation and curves as in Fig. (a). ; (c) $3^-$ inelastic distribution of target. The calculations follow the same color representation and curves as in Fig. (a).}}
\par\end{center}
\label{60ni} 
\end{figure*}

For deforming the imaginary part of the optical nuclear potential, the same prescription adopted for the deformation of its real part is adopted. This means that a multipole expansion of the optical potential is performed considering up to the octupole term. The radial part of the form factor are the derivatives of the potential. So, the transition operator is given by a polynomial  expansion where each term is composed by the product of the deformation parameter, the radial derivative and the spherical function. Due to the large deformation parameter of the $^{16}$O excited state, an option in the FRESCO code guarantees the volume conservation up to second-order correction.

For the $^{60}$Ni nucleus, the excited states are described within
the context of the vibrational model. In the calculations, the reduced
transition probabilities $B(E2)\uparrow=0.093$ e$^{2}$b$^{2}$ for
quadrupole state and $B(E3)\uparrow=0.021$ e$^{2}$b$^{3}$ for the
octupole state, are those reported in Refs. \citep{raman} and \citep{kibedi},
respectively. The $B(E3)\uparrow=0.0015$ e$^{2}$b$^{3}$ for the
$3^{-}$ excited state of $^{16}$O projectile was taken from Ref.
\citep{kibedi}. Since this is the only state of the projectile coupled
in the calculations, there is no difference when considering a vibrational
or a rotational model for the excitation.

 In all
calculations, the finite diffuseness of nuclear surfaces (discussed
in Ref. \citep{brett}) was considered. The respective $\delta$ deformation
lengths and the reduced matrix elements used in the calculations were
obtained from these experimental values using expressions reported
in Ref. \citep{fresco}. Calculations coupling the inelastic excitations of target and projectile are represented, from now on, by a solid blue line. Finally, calculations considering all the above mentioned couplings but disregarding the deformation of imaginary part of nuclear potential will be presented as a red dotted line.

The inclusion of the $2^{+}$ and $3^{-}$ inelastic states of the target $^{60}$Ni
in the CC calculations was not sufficient to reproduce
the elastic scattering at large angles (Fig. $2.a$). Several calculations
were performed trying to obtain a proper imaginary normalization that reproduces data, however, this procedure
did not improve the agreement between the experimental data and the
theoretical elastic angular distributions (especially at large angles). The inclusion of the projectile excitation
in the calculations (solid blue line), using the SPP optical potential,
gives a remarkable good description of elastic scattering
(Fig. $2.a$). The calculation considering all the previous excitations
and no deformation in the imaginary nuclear potential (red dotted line) does not describe the data properly, indicating
the necessity of deforming the imaginary term.

Figs. $2.b$ and $2.c$ show the inelastic angular distributions
of $^{60}$Ni $2^{+}$ and $3^{-}$ states, respectively. CC calculations
including the already mentioned states (solid blue lines) show good
agreement with data. Ref. \citep{jouni} shows that the $3^{-}$ state
of $^{16}$O can be interpreted as the isoscalar octupolar giant mode.
The relevance of such state for a successful data analysis may indicate
the importance of GR in elastic and quasielastic scattering.

We emphasize once again that the the numerical factor  $0.8$ multiplying the imaginary part of the bare potential in our CC
calculations accounts for the high energy of the explored reaction.
As many channels are open at such energy, it is necessary to take
into account the loss of flux to all processes that have not been
explicitly considered. This effect is particularly important here
since we are exploring the internal region of the reacting system where the flux dissipation effect is stronger.

\subsection{The $^{16}$O$+^{27}$Al System}

The same calculation approach can also describe the elastic and
inelastic angular distributions measured in Ref. \citep{linares}
for $^{16}$O$+^{27}$Al reaction at $280$ MeV. In the $^{16}$O$+^{27}$Al
spectrum (see Fig. 1 in Ref. \citep{linares}), it was possible to
distinguish the elastic channel from a composition of five excited states of
target: $1/2^{+}$ ($E^{*}=0.84$ MeV), $3/2^{+}$ ($E^{*}=1.01$
MeV), $7/2^{+}$ ($E^{*}=2.21$ MeV), $5/2^{+}$ ($E^{*}=2.73$ MeV)
and $9/2^{+}$ ($E^{*}=3.00$ MeV). The respective reduced
transition probabilities $B(E2)\uparrow$ values are $0.004$, $0.019$, $0.004$, $0.007$ and $0.004$ e$^{2}$b$^{2}$. Again, a large structure at $\sim6$ MeV containing excited states of the projectile and target has been observed.
The $B(E2)\uparrow$ of $^{27}$Al multiplet states have already been
experimentally determined and can be found in Ref. \citep{aluminio}.
Figs. $3.a$ and $3.b$ show angular distributions for the $^{16}$O$+^{27}$Al
elastic and inelastic scatterings, respectively. 

We reiterate that the bare optical potential calculations and the
CC calculations with the inclusion of the already mentioned target excited
states are not sufficient to describe data. The inclusion of
the $3^{-}$ projectile excitation in the CC calculations is mandatory
for a successful description of both, elastic and inelastic, scattering
angular distributions. This suggests that taking into account the coupling to the $3^{-}$ state of the projectile $^{16}$O is important at energies well above the barrier, however, further studies must be performed.

In particular, without the inclusion of the
projectile excitation and the deformation of the imaginary part of optical potential, it was not possible to find the right phase of the oscillation pattern of elastic scattering (inset of Fig. $3.a$). Without the deformation of the imaginary nuclear potential, the
data are not well reproduced (red dotted line).

In Fig. $3.b$, the experimental cross section is the sum of the quintuplet of
inelastic channels. The ones that most contribute are the $1/2^{+}$
and $9/2^{+}$ (the latter at the largest angles). The experimental angular oscillations on the cross sections
(as well as its absolute value) are well reproduced. 

There are several differences in the present calculations with those performed in Ref. \citep{linares}. In the previous calculations, the excited states of $^{27}$Al were treated as a hole in the $2^{+}$ state of $^{28}$Si. A poor description of elastic scattering was achieved in this framework (Fig. $3.a$ of Ref. \citep{linares}). Also on those calculations, the proton and $\alpha$  transfer channels were considered, however, no satisfactory description of elastic channel was found (Fig. $3.c$ of Ref. \citep{linares}). A good description (Fig. $4.a$ of Ref. \citep{linares}) of data up to $\theta_{c.m.}=40^{\circ}$ was found when the normalization of the real part of nuclear potential was changed to a $0.6$ factor, however, no explanation was found for it. Even on this previous work, it was said that a great effect due the ressonant states could be the reason of the discrepancies between calculations and data. The main difference between that work and the present one is the deformation of the imaginary part of nuclear optical potential to take into account effectively possible couplings to collective states not included explicitly, as the coupling of the $3^{-}$ state of the projectile, which also revealed to be extremely important in calculations. One must also observe that the present calculation describes the backward scattering data ($\theta_{c.m.}>40^{\circ}$) better than the previous one.

\begin{figure}[H]
\begin{centering}
\includegraphics[width=0.45\textwidth]{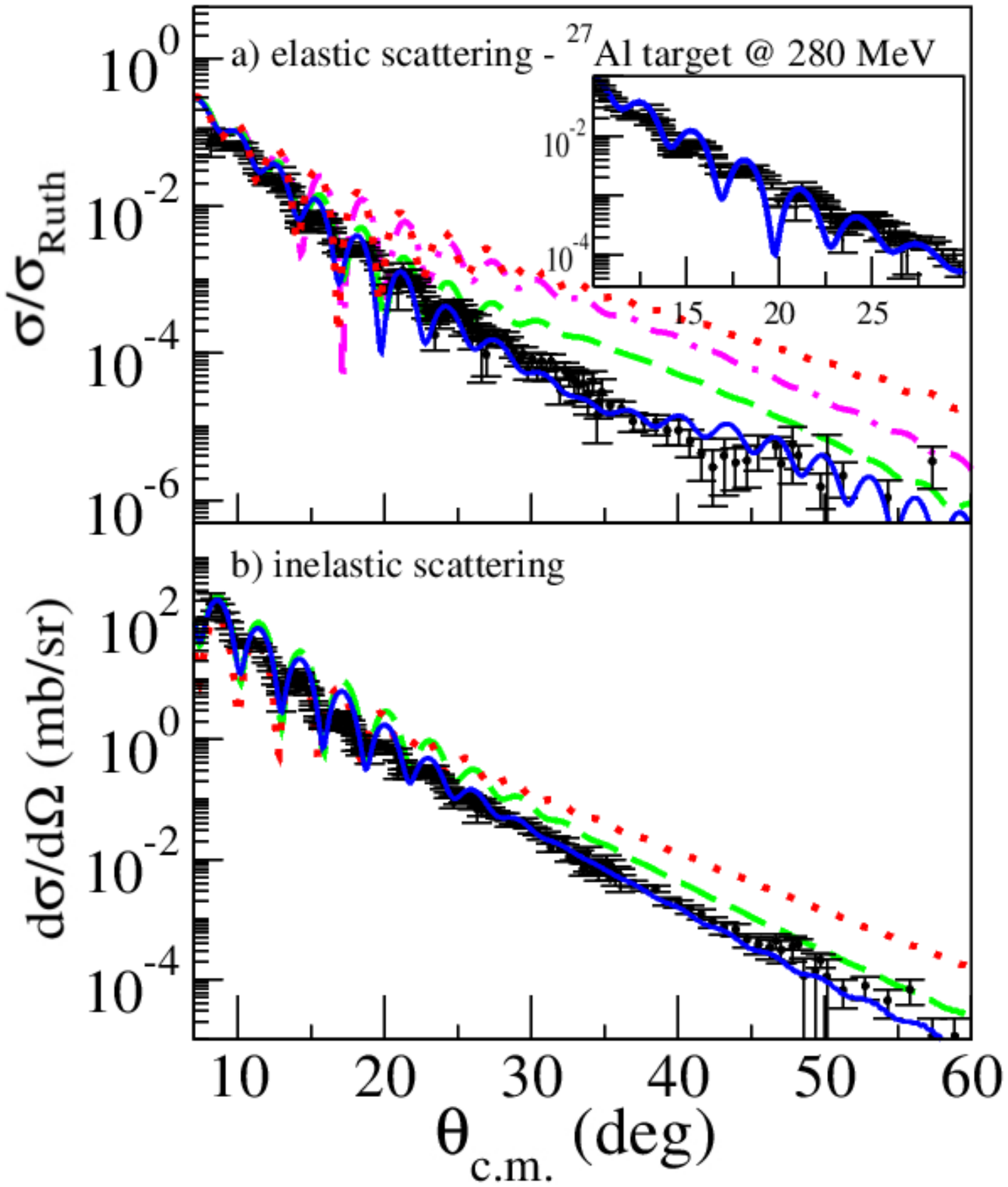} \caption{\textsl{\small{}{}(Color online) - Angular distributions of $^{16}$O$+^{27}$Al at $280$ MeV: (a) Elastic scattering. The dotted-dashed magenta line shows the optical model calculation with the nuclear S\~ao Paulo potential. The dashed green line shows the CC calculations including the inelastic states of the target (see text for details). The solid blue line shows the results of CC calculation including the target excitation and also the first inelastic state of projectile. The dotted red line shows the calculations including all the mentioned couplings but disregarding the deformation of the imaginary potential; (b) Inelastic states of target. The lines represent the summed total inelastic cross section. The calculations follow the same color representation and curves as in Fig. (a).}}
\par\end{centering}
\label{27Al} 
\end{figure}

\section{Discussion \label{sec:discussion}}

In the calculations of $^{16}$O$+^{27}$Al reaction at 280 MeV, the $(1+0.8\cdot i)V_{SPP}$
nuclear potential was utilized (same imaginary normalization of 0.8 as that  of the $^{60}$Ni case). At lower energies one expects that the numerical factor 0.8 to become smaller. This statement was tested by performing the same type of calculations for the $^{16}$O$+^{27}$Al reaction at
$100$ MeV, whose data are presented in Ref. \citep{pereira}. The comparison of data from Ref. \citep{pereira} and the present calculation is presented in Fig. $4$. 

For the lower energy, a good description of elastic (Fig. $4.a$) and inelastic (Fig. $4.b$) channels
were obtained using a $0.45$ normalization for the imaginary part. The decrease in normalization was expected since, at this energy, the number of open channels in the reaction is smaller when compared to the higher one. The comparison of the present calculation with the one presented in Ref. \citep{pereira} demonstrates the importance of considering the coupling of projectile excitation and the deformation of imaginary potential, since they were not considered previously. In the previous work, it was not possible to reproduce the phase of oscillations in elastic scattering (inset of Fig. $2$ of the previous work). This could be achieved in the present paper, as demonstrated in the inset of Fig. $4.a$.     

\begin{figure}[H]
\begin{centering}
\includegraphics[width=0.45\textwidth]{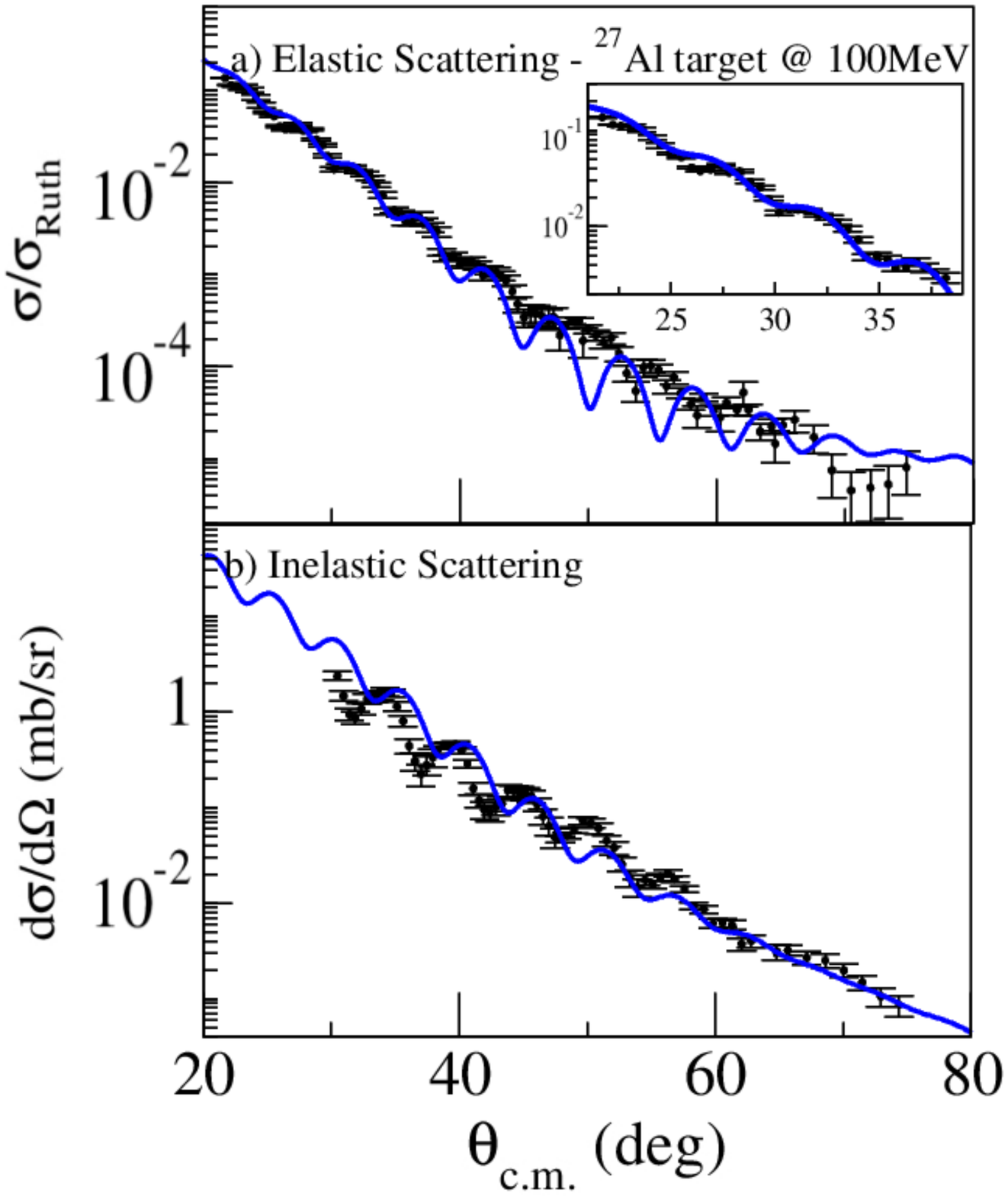} \caption{\textsl{\small{}{}(Color online) - Angular distributions of $^{16}$O$+^{27}$Al at $100$ MeV: (a) Elastic scattering. The solid blue line shows the results of CC calculation including the target excitation and also the first inelastic state of projectile; (b) Inelastic states of target. The lines represent the summed total inelastic cross section.}}
\par\end{centering}
\label{100mev} 
\end{figure}

The present calculations also reproduced the data of the elastic scattering of $^{16}$O$+^{60}$Ni at $42$ MeV incident beam energy (from Ref. \citep{camacho}), using a $40$ times smaller imaginary potential ($V_{bare}$ with $0.02$ normalization of the imaginary part), as would be expected for energies close to the Coulomb barrier. The adopted procedure describes remarkably well data for different systems in an extensive energy range, signalling that the main physics has been considered. In this sense,
it offers a promising approach to analyze heavy-ion reactions well
above the Coulomb barrier, when the couplings of the low-lying
collective states can be strongly influenced by the couplings to very collective states in
the continuum. Under these conditions, due to the large number of open
channels and the possibility to access high excitation energies and
multipolarities, the overall couplings can be accounted for in the calculations
by deforming the imaginary part of nuclear potential.

\begin{figure}[H]
\begin{centering}
\includegraphics[width=0.4 \textwidth]{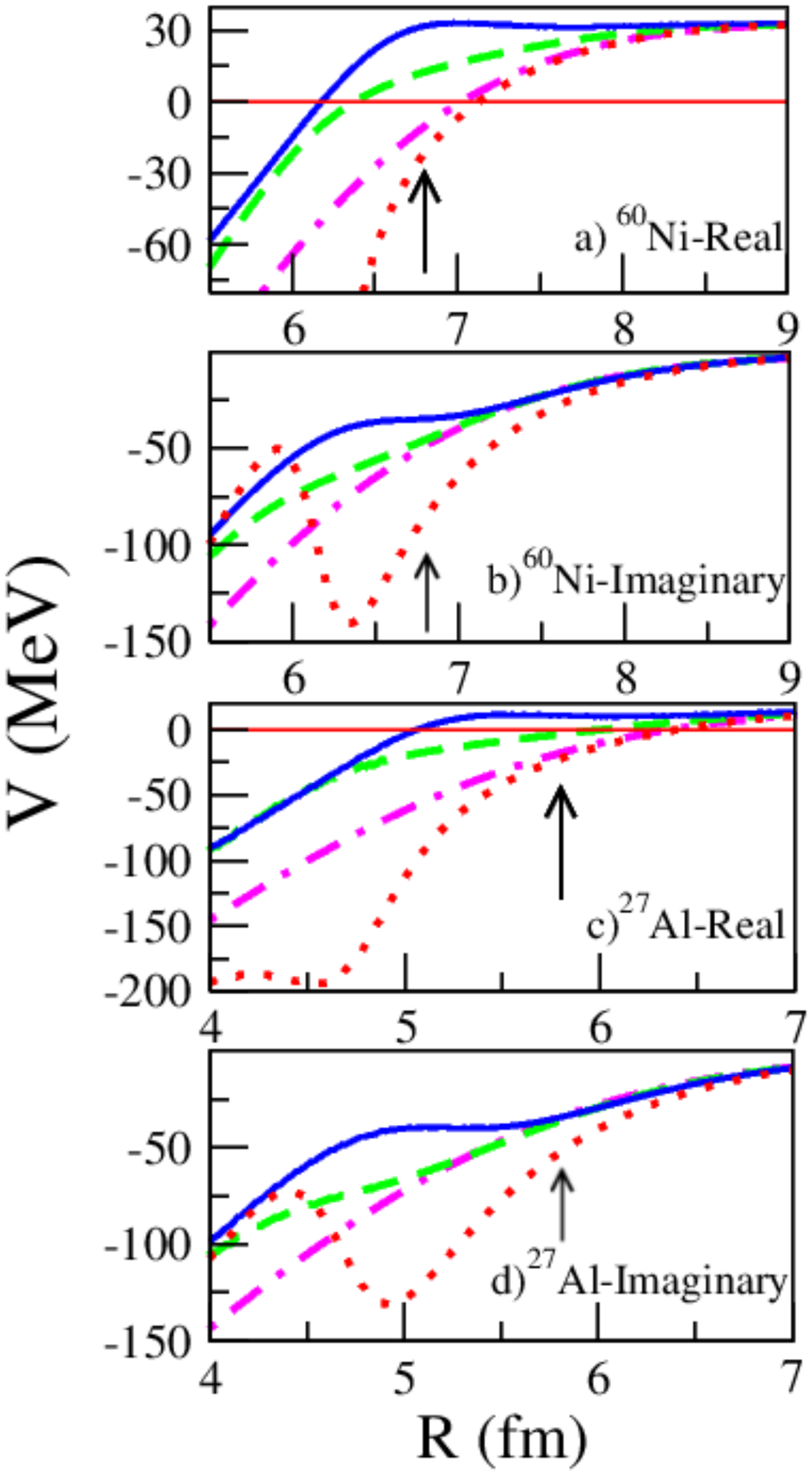} \caption{\textsl{\small{}{}Color online - Effective potentials. (a) Real polarization potential shape for several calculations of $^{16}$O$+^{60}$Ni reaction; (b) Imaginary polarization potential shape for several calculations of $^{16}$O$+^{60}$Ni reaction; (c) Imaginary polarization potential shape for several calculations of $^{16}$O$+^{27}$Al reaction; (d) Imaginary polarization potential shape for several calculations of $^{16}$O$+^{27}$Al reaction. The arrows represent the approximate potential radius position. The line type and color and color code is the same as in Figs. $2$ and $3$.}}
\par\end{centering}
\label{pot} 
\end{figure}

An analysis of the resulting potentials was performed. They were composed by the Coulomb and nuclear interactions, added by the Trivially Equivalent Local Potential (TELP) \cite{telp} implemented in FRESCO code for each individual coupled channel. It is important to mention that one chaneel calculations that includes the TELP have been proved to give a quite reasonable description of the CC elastic angular distribution, even in the case of system involving weakly bound nuclei \cite{rangel}. For the qualitative analysis here proposed it is sufficient to use such approximation. In Fig. $5$ the effective potentials are presented for the real and imaginary parts. For the real part, the Coulomb potential is also included. Figs. $5.a$ and $5.b$ show the real and imaginary parts (respectively) of the total potential for the $^{16}$O$+^{60}$Ni system. Figs. $5.c$ and $5.d$ show the same potentials for the $^{16}$O$+^{27}$Al reaction (exclusively at $280$ MeV). 

An analysis of Figs. $5.a$ and $5.c$ shows that the effect of coupling the inelastic channels of the target and projectile (deforming the imaginary potential) affects the inner part of the barrier, especially changing the region close to the surface, defined, as usual, to be around the potential radius (indicated by the black arrows). The changes in the real part of the total potential accompany the changes (in the same region) of the imaginary part of the optical potential, as can be seen in Figs. $5.b$ and $5.d$. Both imaginary potentials became less absorptive, thus enhancing nuclear transparency in the overlapping region of the colliding nuclei. From Figs. $5.a$ and $5.c$ it is observed that the effect of the real DPP is to slightly widen the Coulomb barrier. Accordingly, considering the DPP in an one-effective channel description, one can assert that such an added potential to the bare one, induces a repulsive real and a weakening of the imaginary components. This is opposite to what happens in the surface region at lower energies, or when one disregards the deformation of the imaginary part (check the red-dotted line in Fig. $5$). This may indicate the necessity of including even the breakup channel since here the energy is quite high and such coupling may be relevant. Further work is required to elucidate the matter.

In general, it is not easy to guess what is the expected effect of the polarization potentials on the elastic and inelastic angular distributions. The real and imaginary parts may have different effects on the elastic and inelastic scattering depending whether they are attractive or repulsive. From Fig. $2.a$, it is observed that coupling the inelastic channels of the target decreases the elastic differential cross sections of the $^{16}$O$+^{60}$Ni reaction at backward angles. Then, coupling the inelastic channel of the projectile (blue solid line) affects especially the angular region between $\theta_{c.m.}=30^{\circ}$ and $\theta_{c.m.}=45^{\circ}$, increasing its differential cross section. For the $^{16}$O$+^{27}$Al system (Fig. $3.a$) occurs the opposite. As already mentioned, the net effect of the polarization potentials depends on the relative importance of the real and imaginary parts of total potential. For the $^{16}$O$+^{60}$Ni case, the polarization potential acted close to the surface, affecting a small region of scattering angles. One should note that, to properly describe data, changes must occur only at  particular values of the radius positions of the polarization potential. These local changes indicate that it is not possible, for instance, to simply find a normalization of the imaginary part that generates an effective potential which would describe the scattering in the the whole angular distribution. Fig. $5$ shows, for both systems, two aspects: $1.$ the necessity to use the deformation of the imaginary part of potential to generate an exclusively repulsive polarization potential; $2.$ the important role played by the projectile excitation, since the coupling to the $3^{-}$ state of $^{16}$O is the one that produces more changes in the shape of the surface of the potential, being the responsible for the changes in the elastic scattering angular distributions. However, one must remember that both effects must be coupled to give a proper description of the angular distribution. 

\section{Conclusions \label{sec:Conclusions}}

Summarising, we have successfully  described the elastic and inelastic scattering of $^{16}$O + $^{60}$Ni and $^{16}$O + $^{27}$Al at intermediate energies, through a coupled channels calculation
involving a deformed optical potential operator. We demonstrated that the deformed imaginary part of the potential, together with the inclusion of the excited $3^{-}$ state of $^{16}$O were important to get a good description of the data. In particular, the role played by $3^{-}$ state of $^{16}$O projectile in the calculations was found to be of paramount importance. New perspectives are open to
extend this exploration to other projectile/target systems and at other incident energies. In addition, an accurate investigation of the effect of coupling to the continuum, accessible with $^{16}$O scattering at intermediate energy, can provide precious information on similar effects in weakly bound nuclei. In fact, in these latter systems, one expect a strong coupling to continuum even at energies around the Coulomb barrier. The precise description of heavy-ion elastic scattering can play a major role for a precise determination of nuclear matrix elements to be extracted from reaction studies. Pertinent examples are the recent studies of heavy-ion charge exchange reactions and their connection with neutrinoless double beta decay \cite{f1,f2}   

\section*{Acknowledgements}

This work was partially supported by INFN (Italy) and FAPERJ, CNPq (Proc. No. $464898$/$2014-5$), FAPESP and
CAPES (Brazil). MSH acknowledges a Senior Visiting Professorship granted by the Coordena\c c\~ao de Aperfei\c coamento de Pessoal de N\'ivel Superior (CAPES), through the CAPES/ITA-PVS program. 
We would also like to thank the technical staff of INFN-LNS for assisting in the maintenance and operation of
the accelerator.

\end{document}